\documentclass[aps,twocolumn,pra]{revtex4-2}

\usepackage[T1]{fontenc}
\usepackage{dsfont}
\usepackage{amsfonts}
\usepackage{amsmath}
\usepackage{bm}
\usepackage{amssymb}
\usepackage{graphicx}
\usepackage{bm}%
\usepackage{braket}

\usepackage[colorlinks=true,urlcolor=blue,citecolor=blue,linkcolor=blue]{hyperref}
\DeclareGraphicsExtensions{.eps,.ps,.pdf}
\newcommand{\diff}{\mathop{}\!\mathrm{d}}
\newcommand{\al}{\alpha}

\newcommand{\la}{\lambda}
\newcommand{\ka}{\kappa}

\newcommand{\La}{\Lambda}

\newcommand{\BM}{\begin{displaymath}}

\newcommand{\EM}{\end{displaymath}}

\newcommand{\ie}{\hbox{\em i.e.{}}}

\newcommand{\eg}{\hbox{\em e.g.{}}}
\def \be  {\begin{equation}}
\def \eeq {\end{equation}}
\def \baa {\begin{eqnarray*}}
\def \eaa {\end{eqnarray*}}
\def \bb  {\begin{thebibliography}}
\def \eb  {\end{thebibliography}}

\def \lab #1 {\label{#1}}

\def \Tr {\text{Tr}}

\newcommand{\rhs}{\hbox{r.h.s.{}}}
\newcommand{\Rr}{\mathsf{R}}

\newcommand{\da}{\dagger}

\makeatletter
\g@addto@macro\bfseries{\boldmath}
\makeatother

\usepackage[dvipsnames]{xcolor}

\usepackage[normalem]{ulem}
\usepackage{xcolor}
\usepackage{soul}
 
 \newcommand\redout{\bgroup\markoverwith
{\textcolor{red}{\rule[.5ex]{2pt}{2.4pt}}}\ULon}
 \newcommand\blueout{\bgroup\markoverwith
{\textcolor{blue}{\rule[.5ex]{2pt}{2.4pt}}}\ULon}

\begin{document}
\widetext

\title{Metastable spin-phase diagrams in antiferromagnetic Bose-Einstein condensates}

\author{E.{} Serrano-Ens\'astiga}
\email{edensastiga@ens.cnyn.unam.mx}
\affiliation{Departamento de Física, Centro de Nanociencias y Nanotecnología, Universidad Nacional Autónoma de México\\
Apartado Postal 14, 22800, Ensenada, Baja California, México}

\author{F.{} Mireles}
\email{fmireles@cnyn.unam.mx}
\affiliation{Departamento de Física, Centro de Nanociencias y Nanotecnología, Universidad Nacional Autónoma de México\\
Apartado Postal 14, 22800, Ensenada, Baja California, México}

\begin{abstract}
Spinor Bose-Einstein condensates under external magnetic fields exhibit well-characterized spin domains of its ground state due to spin-dependent interactions. At low temperatures, collision-induced spin-mixing instabilities may promote the condensate to dwell into metastable states occurring near the phase boundaries.  
In this work, we study theoretically the metastable spin-phase diagram of a spin-1 antiferromagnetic Bose-Einstein condensate at zero and finite temperatures. The approach makes use of Hartree-Fock theory and exploits the symmetry of the Hamiltonian and of the order parameters yielding a closed system of transcendental equations for the free energy, fully avoiding the use of selfconsistency. Our results are consistent with recent experiments and allow us to explain qualitatively the different types of observed quench dynamics. In addition, we found that similar phenomena should occur in antiferromagnetic spinor condensates with a sudden change in the temperature. It is shown also that the increase of temperature induces a traceable shift of the Ferromagnetic-Polar transition boundary, behavior previously not noticed by selfconsistent mean-field calculations.
\end{abstract}
\maketitle
\section{Introduction}
\label{sec.Int}
Spinor Bose-Einstein condensates (BECs) of ultracold atoms can be manipulated nowadays with astonishing precision offering unprecedented opportunities to study spin-dependent many-body physics.\cite{Kaw.Ued:12,pethick2008bose,lewenstein2012ultracold} Of crucial importance is the underlying physics of the phase diagram in spinor BECs, where the nature of the spin domain phases strongly depends on the atomic species and on the external fields. The study of the spin-phase diagram in spinor BECs via mean-field theories were introduced first for spin $f=1$ \cite{PhysRevLett.81.742,ohmi1998bose} and subsequently for higher spins  ($f=2,3,4,6,8) $\cite{Kaw.Ued:12,ciobanu2000phase,Bar.Tur.Dem:06,diener2006cr,PhysRevA.84.053616}, where the usual parameters of the phase diagram are the coupling factors of the different spin-dependent interactions, and/or the coupling factors of the linear and quadratic Zeeman interactions. The predicted phase diagrams have been confirmed for spin-1 BECs with antiferromagnetic ($^{23}$Na \cite{stenger1998spin,stamper.Andrews.etal:1998,Jacob:12}) and ferromagnetic spin-dependent interactions ($^{87}$Rb \cite{Chang.Hamley.etal:2004} and $^7$Li  \cite{PhysRev.2.033471}), to mention a few.

It is known that in spinor BECs there may occur the coexistence of several domain phases in which its different spin states and associated order parameters are not continuously transformed at the phase boundary leading to a first-order transition. Owing the non-continuity of the order-parameters, both phases may remain stable, whereas the ground state is given by the lowest energy state. Consequently, near the phase boundaries may exist metastable phases \cite{PhysRevLett.78.3594,PhysRevA.61.063613,PhysRevA.74.033612,PhysRevA.78.023632,PhysRevA.73.013629,PhysRevA.88.043629,jimenez2019spontaneous} which play a crucial role in a variety of phenomena such as quantum tunneling \cite{sta.mie.chi:1999,PhysRevLett.82.2228}, domain formations \cite{PRL.110.165301,NJP.095003}, and quench dynamics \cite{PRA.98.063618,PhysRevA.99.023606,PRA.100.013622}, among others \cite{PRA.90.023610,PRA.95.053638}. In particular, recent experiments have reported the observation of dynamical quantum phase transitions under different types of quench dynamics in an antiferromagnetic spinor BEC \cite{PRA.100.013622}, including a more recent experiment that involves a phase transition between excited states \cite{PRL.124.043001}. A dynamical phase transition refers to a non-analytical  change in the quench dynamics of the condensate \cite{RPP.81.054001}, occurring by the sudden change of the controlling parameters that modifies the free energy and stability of each spin phase. Under this perspective, the different quench processes and the existence of dynamical phase transitions can also be understood through the analysis of its corresponding metastable spin-phase diagram.

In this work, we analyze within Hartree-Fock (HF) theory the emergence of  metastable spin domains in a spin-1 antiferromagnetic condensate at finite temperatures. The resulting phase diagrams offer further insights of the nature of the different quench processes observed in
\cite{PRA.100.013622}.  
Our approach starts with the Hartree-Fock (HF) approximation \cite{blaizot1986quantum,griffin2009bose,Kawa.Phuc.Blakie:2012}
but take advantage of the common symmetries between the Hamiltonian and the order parameter, to then reduce the problem to the solution of a system of algebraic-transcendental equations instead to appeal for self-consistency. This framework leads us to closed expressions for the study of the metastable phase diagrams and their physical properties, including analytical expressions at low temperatures. The formalism is quite general and can be applied for a spinor condensate of any spin value and any spin-dependent interaction in mean-field theory.

We show the appearance of overlapping regions of the spin domains which tend to increase as the temperature is increased. This allows us to infer similar quench processes due to a sudden change of the temperature, instead of an abrupt change in an external field as done experimentally in \cite{PRA.100.013622}. We characterize the spin phases in the overlapping regions through calculations of the spin magnetization and the atom fraction in the magnetic sublevels in order to further distinguish the spin domains among each other. We also uncover a sizable shift of the Ferromagnetic-Polar (FM-P)  phase boundary driven by temperature,  in sharp contrast with earlier selfconsistent HF results that predicts a fix phase boundary with temperature \cite{Kawa.Phuc.Blakie:2012}. Moreover, the approach enables us to extract a useful analytical approximation of the FM-P phase boundary valid for a wide range of temperatures.

%
%
%
\section{Model} 
\label{sec.Mod}
Let us consider a dilute $f=1$ spinor Bose-Einstein gas confined in an optical trap with potential $U(\bm{r})$, and subject to linear ($p$) and quadratic ($q$) Zeeman fields oriented along the $z$ axis. The coupling factors $(q,p)$ can be manipulated independently in laboratory \cite{Ger.Wid.Blo:06,PhysRevLett.107.195306} and here will be our parameters in the phase diagram. We restrict ourselves to $p\geq 0$ values because the phase diagram is symmetric by inversion $p\rightarrow -p$ \cite{Kaw.Ued:12}. The system is assumed to be weakly interacting and sufficiently diluted such that only two-body collisions are predominant and the $s$-wave approximation is still valid. The full Hamiltonian of the atomic gas, constituted by the single-particle and interaction terms, is written in the second-quantization formalism as \cite{Kaw.Ued:12,lewenstein2012ultracold}
\begin{align}
& \hat{H} 
= \int \diff \bm{r} \Big\{  \hat{\bm{\Psi}}^{\da} 
\left( h_{s} \mathds{1}_3
 - p F_z + q F_z^2 
\right) \hat{\bm{\Psi}} 
\label{Full.Ham}
\\
& + \frac{c_0}{2} \sum_{i,j}
 \hat{\psi}_i^{\da} \hat{\psi}_j^{\da} \hat{\psi}_j \hat{\psi}_i
 + \frac{c_1}{2} \sum_{\al,i,j,k,l} 
(F_{\al})_{ij} (F_{\al})_{kl} \hat{\psi}_i^{\da} \hat{\psi}_k^{\da} \hat{\psi}_l \hat{\psi}_j
\Big\} \, ,
\nonumber
\end{align}
where $h_{s} = - \hbar^2 \nabla^2/2M  + U(\bm{r})$ is the spatial Hamiltonian.  $\mathds{1}_3$ is the $3\times 3$ identity matrix and $F_{\al}$ are the angular momentum matrices of spin $f=1$ with $\al=x,y$ or $z$.
The spinor-quantum field associated to the spinor condensate is denoted by  $\hat{\bm{\Psi}}= (\hat{\psi}_1 \, , \hat{\psi}_0 \, , \hat{\psi}_{-1})^{\text T}$, where $\hat{\psi}_m$ are the field operators  with $m=-1,0,1$ the possible magnetic quantum numbers, and T denotes the transpose. From now on, we will employ bold Greek symbols for the 1-spinors and bold Latin symbols for the 3-dimensional vectors in real space. The Hamiltonian \eqref{Full.Ham} has a symmetry group isomorphic to $SO(2) \times \mathds{Z}_2 $, constituted by the rotations about the $z$ axis, and the reflection across the $yz$ plane. The spin-independent and spin-dependent coupling factors, $c_0$ and $c_1$, respectively, are related to the $s$-wave scattering lengths $a_0$ and $a_2$ of the total spin-$F$ channel $a_F$ ($F=0,2$) \cite{PhysRevLett.81.742,ohmi1998bose}
\begin{equation}
c_0 = \frac{4\pi \hbar^2(a_0 + 2a_2)}{3M} \, , \quad c_1 = \frac{4\pi \hbar^2(a_2 -a_0)}{3M} \, ,
\end{equation}
where $M$ is the atomic mass. Experimental measurements indicate that $^{23}$Na atoms have $a_0=47.36(80)a_B$ and $a_2=52.98(40)a_B$ \cite{Kaw.Ued:12,PhysRevA.63.012710}, with $a_B$ the Bohr radius, yielding a BEC with antiferromagnetic interactions ($c_1>0$). For our theoretical calculations, we consider the following values of the coupling factors $c_0/a_B^3 =0.42eV$ and $c_1 =c_0/27$, derived with scattering lengths $a_F$ within the quoted uncertainties. Other species could present ferromagnetic interactions ($c_1<0$) as $^{87}$Rb \cite{Chang.Hamley.etal:2004} and $^7$Li  \cite{PhysRev.2.033471}.

Mean-field approximation at zero temperature ($T=0$) assumes that all the atoms in the spinor condensate are in the same quantum state described by an spinor order-parameter $\langle \hat{\bm{\Psi}} \rangle = \bm{\Phi}$ \cite{Kaw.Ued:12,lewenstein2012ultracold}. The ground state $\bm{\Phi}= (\phi_1 , \phi_0 , \phi_{-1})^{\text{T}}$ of the BEC minimizes the functional mean-field energy $ E[\bm{\Phi}]= \langle\hat{H} \rangle$. We shall consider the uniform case with $U(\bm{r})=0$ and, hence, the ground state $\phi_j (\bm{r}) = \phi_j e^{i  \bm{k}\cdot \bm{r}}$ with $\bm{k} = \bm{0}$. $E[\bm{\Phi}]$ constrained to a fixed number of particles $N= \bm{\Phi}^{\da} \bm{\Phi}$ is reduced to 
\begin{align}
E[\bm{\Phi}] = & \bm{\Phi}^{\da} \left( -p F_z + q F^2_z \right) \bm{\Phi} + \frac{c_0}{2} \left(\bm{\Phi}^{\da} \bm{\Phi} \right)^2
\nonumber
\\
&
 + \frac{c_1}{2} \sum_{\al} \left( \bm{\Phi}^{\da} F_{\al} \bm{\Phi} \right)^2 - \mu \left(  \bm{\Phi}^{\da} \bm{\Phi} -N \right) \, ,
 \label{energy.GS}
\end{align}
where $\mu$ is the chemical potential, {\it i.e.}, the required energy to add an atom to the condensate. The conditions $
 \delta E[\bm{\Phi}]/ \delta \phi_m^* =0 $ 
yield the so-called multi-component Gross-Pitaevskii (GP) equations. For spin-1 we have five solutions of the GP equations \cite{stenger1998spin,Kaw.Ued:12}, identified through the spinor order-parameter $\bm{\Phi}$. Here we only scrutinize the solutions for the antiferromagnetic case $c_1>0$ and $p\geq 0$, which leads to the following phases of the spin-1 BEC:

$\bullet$ Ferromagnetic (FM) phase: The spinor order-parameter is equal to $\bm{\Phi} = \sqrt{N}(1,0,0)^{\text{T}}$. It is symmetric under rotations about the $z$ axis, imposing that the symmetry group is isomorphic to the special orthogonal group $SO(2)$. Each atom is fully magnetized along the $z$ axis, $M_z \equiv \langle F_z \rangle/N = 1$ and $M_{\perp} \equiv (\langle F_x \rangle^2 + \langle F_y \rangle^2 \rangle )^{1/2}/N = 0$. 

$\bullet$ Polar (P) phase: Here $\bm{\Phi}= \sqrt{N}(0,1,0)^{\text{T}}$. Its symmetry group, which is isomorphic to $SO(2) \times \mathds{Z}_2 \cong O(2)$, consists to rotations about the $z$ axis and time-reversal symmetry (equivalent to an inversion through the origin) \cite{Bengtsson17}. It has zero magnetization $M_z = M_{\perp} = 0$. 

$\bullet$ Antiferromagnetic (AF) phase: It consists of a family of quantum states $\bm{\Phi} = \sqrt{N}( \cos \chi , 0, \sin \chi)^{\text{T}}$ with $\chi \in (0,\pi/4]$. A family of states that represents a phase are also called non-inert states \cite{Mak.Suo:07}. The whole set is symmetric over two geometric operations, a rotation by $\pi$ about the $z$ axis, and a reflection across the $yz$ plane), implying that the symmetry group is isomorphic to $\mathds{Z}_2 \times \mathds{Z}_2$. Notice that the AF phase tends to the FM phase when $\chi \rightarrow 0$. On the other hand, the AF phase tends to the P phase but oriented over the $y$ axis when $\chi \rightarrow \pi/4$. The magnetic moment depends on $\chi$, $M_z = \cos(2\chi)$ and $M_{\perp}=0$.

\begin{figure}[t!]
 \scalebox{0.6}{\includegraphics{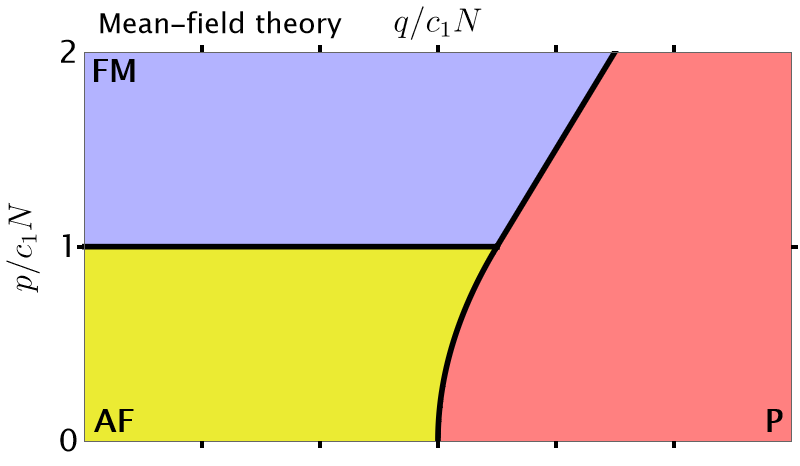}}

 \scalebox{0.6}{\includegraphics{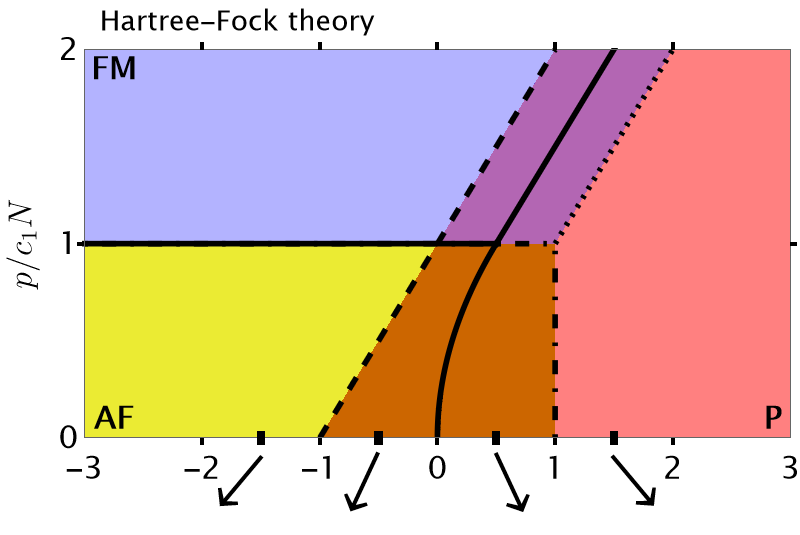}}
  
\vspace{-0.1cm} \scalebox{0.255}{\includegraphics{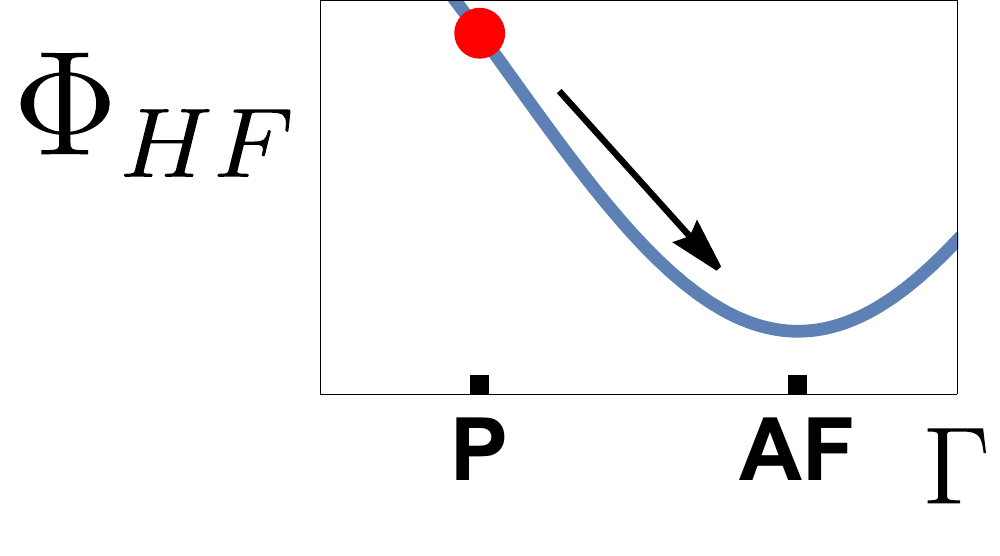}}
\scalebox{0.18}{\includegraphics{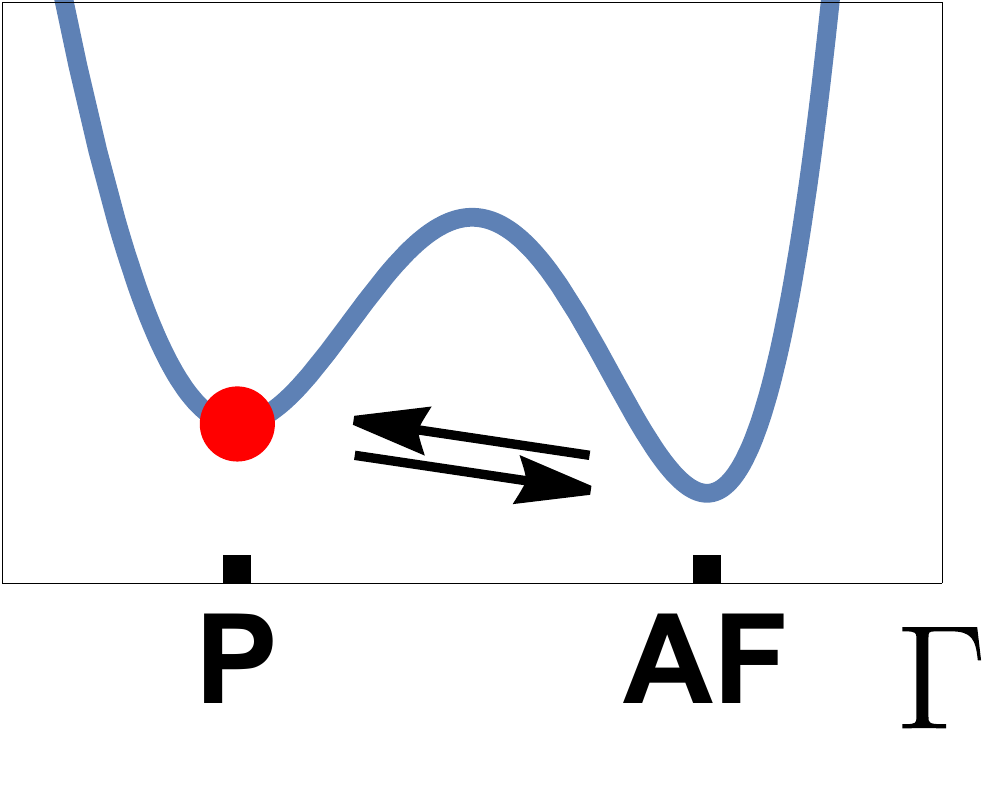}}
\scalebox{0.18}{\includegraphics{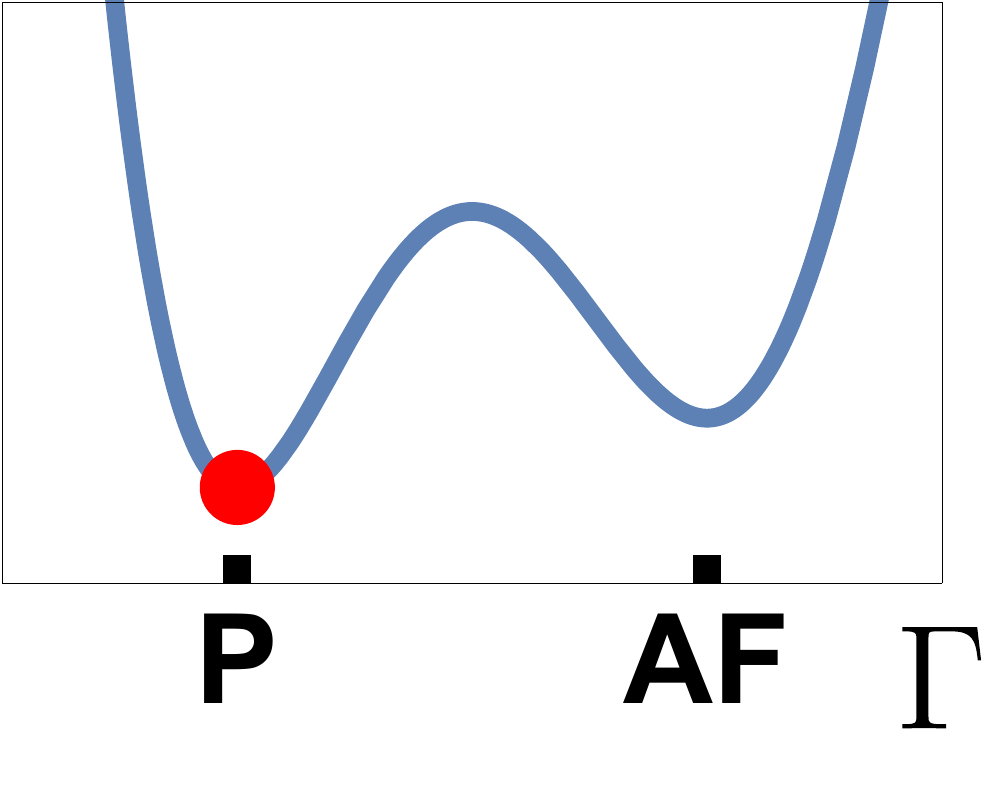}} 
\scalebox{0.18}{\includegraphics{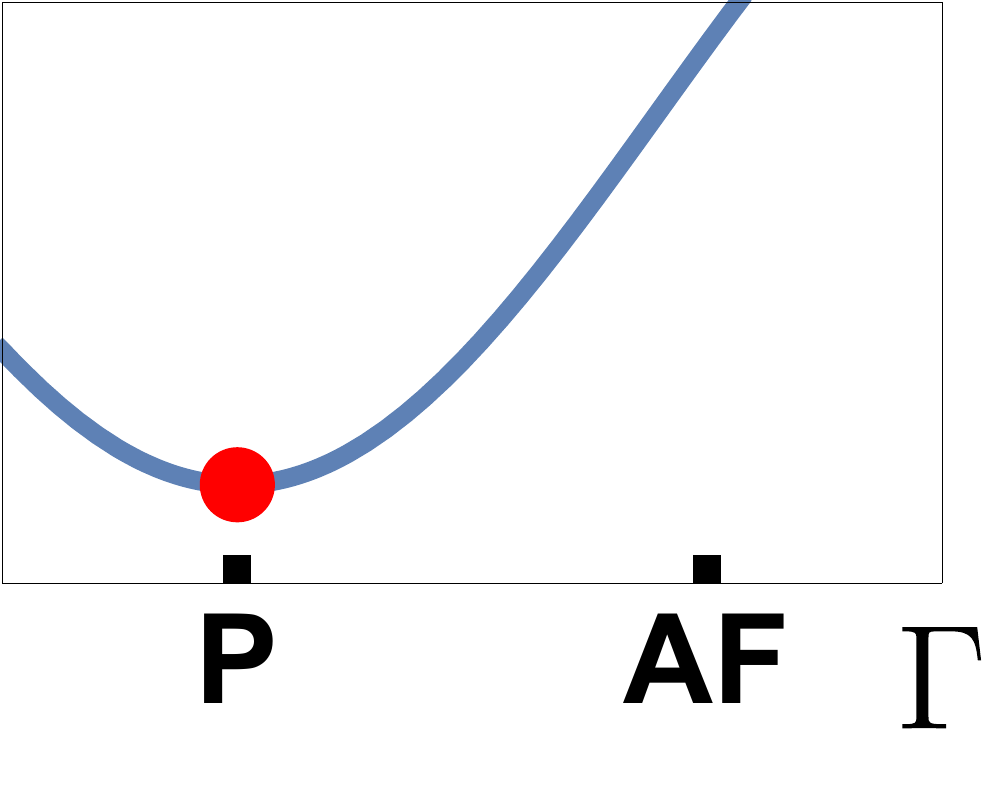}}
\caption{\label{Fig1} (Color online) \emph{Top}: Spin-phase diagram of the spin-1 BEC gas at $T=0$ calculated with the mean-field approximation \cite{Kaw.Ued:12}. The FM, P and AF phases are denoted by blue, red, and yellow, respectively. The solid lines define the phase transitions. \emph{Center}: Metastable spin-phase diagram of the spinor condensate at $T=0$ calculated with the HF theory and the approach described in Sec.~\ref{sec.met}. The overlaps between the phases are denoted by their respective secondary color. The boundaries of the allowed regions are indicated by dotted (FM), dashed (P), and dashed-dotted (AF) lines, respectively. \emph{Bottom}: Schematic graphics of the thermodynamic potential $\Phi_{HF}$ versus the order-parameter variables of $\rho^c$ and $\rho^{nc}$, encompassed in $\Gamma$, for $(q,p)=(n,0)c_1N$ with $n=-1.5,-0.5,0.5,1.5$, respectively.  The red point indicates the initial state of the spinor condensate (P phase) and the arrows denote the phase transition, via quantum tunneling or by the presence of an effective force.}
\end{figure}

The FM and P phases exist, in principle, for each $(q,p)$-values. On the other hand, the AF phase satisfies \cite{Kaw.Ued:12}
\begin{equation}
\label{cond.AFM1}
\cos \chi= \sqrt{\frac{c_1 N+p}{2c_1 N}} \, ,
\end{equation}
which implies that the AF phase does not exist for $p>c_1 N$. The limit case $p=c_1 N$ saturates the spin magnetization of the AF phase $M_z=1$. The phase diagram in the case for zero temperature ($T=0$) is amenable to analytic results, plotted in Fig.~\ref{Fig1} \cite{stenger1998spin}. The ground state for each $(q,p)$ point is the one that minimizes \eqref{energy.GS}. The solid lines delimit the phases.

The Hartree-Fock (HF) approximation \cite{blaizot1986quantum,griffin2009bose,Kawa.Phuc.Blakie:2012} which, despite of being the simplest many-body theory after the mean-field approximation, it is known to capture the main relevant physics occurring in the phase diagram and its concomitant boundary regions. Moreover, it allows us to study the spinor condensates at finite temperatures \footnote{The influence of the temperature on the phase diagram for spinor condensates have been previously investigated theoretically \cite{griffin2009bose,PhysRevA.90.043609,PhysRevA.70.043611,Kawa.Phuc.Blakie:2012,phuc2013beliaev} and experimentally \cite{PhysRevA.70.031602,PhysRevLett.111.025301,PhysRevA.91.033635,PhysRevLett.119.050404}.}. Formally, in the HF approximation the field operator is given by the order parameter and a perturbation $\hat{\delta}_j$, \ie, $\hat{\psi}_j = \phi_j + \hat{\delta}_j$. For simplicity, we neglect the three-field correlations $\langle  \hat{\delta}_i \hat{\delta}_j \hat{\delta}_k^{\da} \rangle$ and the anomalous density $\langle \hat{\delta}_i \hat{\delta}_j \rangle$ (Popov approximation), which gives a reasonable first approximation for diluted gases at all temperatures \cite{Griffin:96}. We now apply the elements of the HF theory following Ref.~\cite{Kawa.Phuc.Blakie:2012} with slight
changes in the notation. However, in contrast with Ref. ~\cite{Kawa.Phuc.Blakie:2012}, we shall avoid the self-consistent procedure as it is explained in Sec.~\ref{sec.met}. 

The condensate (\emph{c}) and non-condensate (\emph{nc}) atoms are represented by a density matrix $\rho^c_{ij} =N^c \phi^*_j \phi_i $ and $ \rho^{nc}_{ij}= \langle \hat{\delta}_j^{\da} \hat{\delta}_i \rangle $, respectively. The trace of each density matrix is equal to the number of atoms of each part, $\Tr (\rho^a) = N^a$ for $a=n, nc$. The non-condensate atoms $\rho^{nc}$ act as a cloud of atoms thermally excited that interacts non-trivially with the condensate fraction $\rho^c$. The total system is then denoted by $\rho =\rho^{c}+\rho^{nc}$ with $\Tr \rho = N = N^c +N^{nc}$. Hence, the HF energy, with its Lagrange multiplier $\mu ( N -\Tr \rho)$, is given by 
 \begin{align}
\label{HF.energy}
&E_{HF} =   E_{s} + \Tr \left[ \rho \left( -p F_z + q F_z^2 \right) \right] -\mu \left( \Tr \rho- N \right)
\nonumber
\\
& + \frac{c_0}{2}\Big( N^2 + \Tr \left[ \rho^{nc} \left(2\rho^c + \rho^{nc} \right) \right] \Big)
\\
& + \frac{c_1}{2} \sum_{\alpha} \Big(  \Tr \left[ \rho F_{\al} \right]^2 
+ \Tr \left[ F_{\al} \rho^{nc} F_{\al} \left( 2 \rho^c + \rho^{nc} \right) \right] \Big) ,
\nonumber
\end{align}
 where the trace involves a summation over the spatial and spinor quantum numbers. For $U(\bm{r})=0$, $E_{s}$ is the kinetic energy, and then, the spatial quantum number is the wavevector $\bm{k}$. The two-body interactions have two terms, the direct and exchange interactions. The effect of each term over the spin coherence and the distribution of the atoms in the magnetic sublevels is discussed in \cite{Kawa.Phuc.Blakie:2012}.

The condensate fraction of the system $\rho^c=N^c \bm{\Phi} \bm{\Phi}^{\da}$ is a pure state with $\bm{k}=\bm{0}$. Hence, the resulting GP equations $\delta E_{HF}/\delta \phi^*_m =0 $ are given by a system of three (non-linear) equations involving $\phi_m$ and $\rho^{nc}$. On the other hand, $\rho^{nc}$ is written as a sum of its eigenvectors $\bm{\xi}^{\la} = (\xi_1^{\la}, \xi_0^{\la}, \xi_{-1}^{\la})^T$ weighted by their Bose-Einstein distribution factor $n_{\la}$, \begin{equation}
\rho^{nc}_{ij} = \sum_{\la} n_{\la} \xi^{\la}_i \xi^{\la*}_j  
\, , \quad
n_{\lambda} =\left( e^{\beta \epsilon_{\la}} -1 \right)^{-1}
\, .
\end{equation}
 The global subindex $\la$ includes the spatial and spinor quantum numbers, $\la= (\bm{k}, \nu)$, with $\nu=1,2,3$ and $\beta= 1/k_{B}T$ where $k_{B}$ is the Bolztmann constant. The eigenvectors $\bm{\xi}^{\la} $ and their associated energies $\epsilon_{\la}$ are obtained by the non-condensate Hamiltonian $A$, given by $A_{ij} = \delta E_{HF}/\delta \rho^{nc}_{ji} $. The decoupling of the spatial and spinor parts in the Hamiltonian $A$ leads to $\epsilon_{\lambda}= -\hbar^2 k^2/2M + \kappa_{\nu}$, with $\kappa_{\nu}$ the eigenvalue of the spinor part of $A$. The spatial part of $\rho^{nc}$ can be integrated using that $\sum_{\bm{k}}\rightarrow (2\pi)^{-3} \int\diff \bm{k}$,
\begin{equation}
\label{Poly.sta}
\rho^{nc}_{ij} = \sum_{\nu=1}^3 \xi^{\nu}_i \xi^{\nu *}_j  \frac{Li_{3/2}\left(e^{-\beta \kappa_{\nu}} \right)}{\la_{dB}^3}  \, ,
\end{equation}
where $Li_{3/2}(z)$ is the polylogarithm function and $\la_{dB} = h / \sqrt{2\pi M k_B T} $ is the thermal de Broglie wavelength. The eigendecomposition of $A_{ij}$, which is now a $3\times 3$ matrix, are called the HF equations.
\section{Approach and results.}
\label{sec.met}
Usually, the GP-HF equations are solved self-consistently \cite{blaizot1986quantum,griffin2009bose,Kawa.Phuc.Blakie:2012}. Here, we start by restricting $\rho^c$ to a particular phase, FM, P or AF. Each phase exhibits some symmetries in common with the full Hamiltonian \eqref{Full.Ham}. By a well known result (Sec. 8.4 of \cite{blaizot1986quantum}), the perturbation of the system $\rho^{nc}$ inherits the common symmetries of $\rho^c$ and $\hat{H}$, reducing its degrees of freedom. This approach simplifies the GP-HF equations from a set of equations of the components of $\rho^c$ and $\rho^{nc}$, to a system of three algebraic-transcendental equations of the $\kappa_{\mu}$ energies. 

Let us illustrate our approach by considering the FM phase, $\bm{\Phi}= N^c(1,0,0)^T$. In general, one must find the variables of $\rho_{ij}^{nc}$ and $N^c$, subject to the condition $N^c + N^{nc}=N$, via self-consistency in the HF theory. However, the degrees of freedom of $\rho^{nc}$ are reduced by its symmetries, consisting of the rotations over the $z$ axis by a generic angle $\theta$, $\Rr_z(\theta) = e^{-i\theta F_z}$, inherited by the FM phase and the Hamiltonian \eqref{Full.Ham}. In particular, this implies that $\rho^{nc}$ must commute with the generator $F_z$, and hence $\rho^{nc}$ and $F_z$ must have the same eigenvectors
\begin{equation}
\rho^{nc}= \sum_{m=-1}^1 \La_m \ket{1,m} \bra{1,m}   \, , \quad \La_m = \frac{Li_{3/2}(e^{-\beta \ka_{m}})}{\la_{dB}^3} \, ,
\end{equation}
where the eigenvalues are given by Eq.~\eqref{Poly.sta}. The atom fractions $N^c$ and $N^{nc}$ can be written in terms of $\ka_m$ because $N^{c}=N-N^{nc}$ and $N^{nc}= \sum_{m} \La_m$, reducing the unknown variables to the three eigenenergies $\ka_{m}$ of the HF Hamiltonian $A= \delta E_{HF} / \delta \rho^{nc}_{ji}$. Moreover, the $A$ matrix, dependent of $\rho^c$ and $\rho^{nc}$, can also be written in terms of the $\Lambda_m$ and then of the $\ka_{m}$ variables. Finally, we use the fact that $A$ and $\rho^{nc}$ share the same eigenvectors, leading us to obtain a system of three algebraic-transcendental equations for the $\ka_m$ eigenenergies,  $\ka_{m}= \bra{1,m} A \ket{1,m}$. The resulting equations for the FM phase and the full procedure for the P and AF phases are described in Appendix \ref{App.B}.

The approach is applied to calculate the allowed region of each phase for several temperatures $T$ in the interval $[0,0.4T_0]$, where $T_0=1.5 \mu K$ is the critical temperature for a scalar BEC with an atomic density $N= 10^{14}cm^{-3}$ \cite{stamper.Andrews.etal:1998}. The results reported in this work  were obtained by solving numerically the equations \eqref{kappa.FM}, Eqs. \eqref{kappa.P}, and the Eqs. \eqref{kappa0.AF}-\eqref{Ener2.AF} for the FM, P, and AF phase, respectively.
\subsection{Metastable phase diagrams}
Let us discuss first the results for $T=0$ \footnote{The numerical solutions for $T=0$ were calculated as the limit $T \rightarrow 0$, considering the values $T=10^{-n}T_0$ with $n=3,\dots,6$.} plotted in Fig.~\ref{Fig1}. We denote the allowed region of each phase with a primary color: the FM, P and AF phases with blue, red and yellow, respectively. Remarkably, we find overlapping regions, implying that there could be some metastable phases upon some $(q,p)$ values. The overlaps are marked with the respective secondary color. The boundaries of the FM, P, and AF phases are denoted with dotted, dashed, and dotted-dashed lines, respectively. The solid lines delimit the region where each phase is the ground state, which minimizes the thermodynamic potential $\Phi_{HF}=E_{HF}-TS_{HF}$, with $S_{HF}$ the HF entropy (\cite{blaizot1986quantum}, see also Appendix \ref{App.A}).
\begin{figure}[t!]
 \scalebox{0.6}{\includegraphics{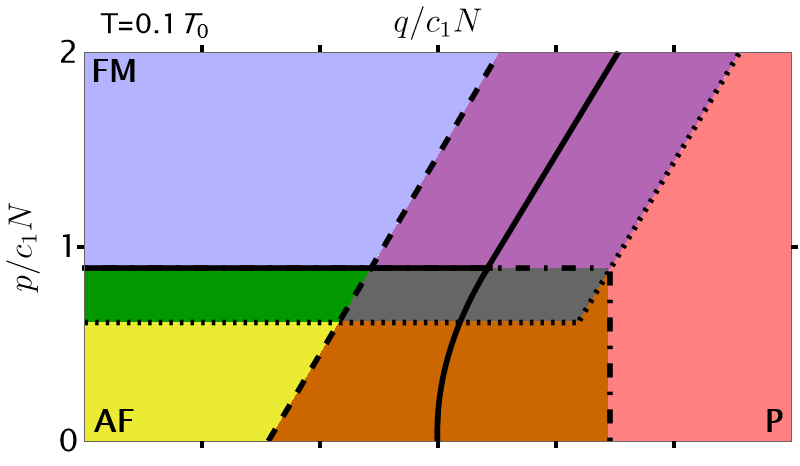}}
 
 \scalebox{0.6}{\includegraphics{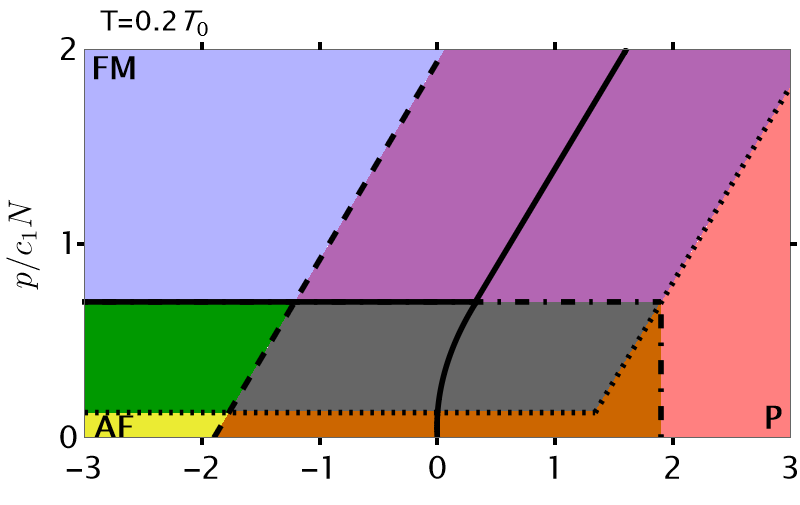}}
\caption{\label{Fig1a} (Color online) Metastable phase diagrams at temperatures $T$ equal to $0.1 T_0 $ (\emph{top}) and b) $0.2T_0$ (\emph{bottom}). Color conventions are as in Fig. \ref{Fig1}. The three phases coexist in the dark-gray region. }
\end{figure}

The quench dynamics of a spinor condensate, \ie, its evolution after the sudden change of a control parameter, could be explained with a metastable phase diagram. For an example at hand, let us consider first that our spinor condensate is prepared in the P phase with control parameters $(q_0,p)=(1.5c_1 N,0)$ at $T=0$. Here the only admissible phase is the polar one, hence the graph of the thermodynamic potential $\Phi_{HF}$ versus the order-parameter variables $\Gamma= \{ \rho^c , \rho^{nc} \}$ has a global minimum (see Fig.~\ref{Fig1}). Essentially, there are three possible $\Phi_{HF}$ with substantial changes to affect the quench dynamics of the spinor condensate. We exemplify each case in Fig.~\ref{Fig1}, corresponding to the points with $p=0$ and the following values of $q$:

$\bullet$ $q_1=0.5c_1 N$: $\Phi_{HF}$ has now two minima related to the P and AF phases. The atoms of the spinor condensate would rather stay in the P phase by the difference $\Delta \Phi_{HF}$. 

$\bullet$ $q_2=-0.5c_1 N$: Here, we also have two admissible phases, but now the AF phase is the ground state. The quantum tunneling is now stimulated to the AF phase. The smaller is $\Delta \Phi_{HF}$ between the phases, more atoms oscillate between the phases with respect to the time. However,  dissipation-energy effects would favor the tendency of the atoms to migrate to the ground state.

$\bullet$ $q_3=-1.5c_1 N$: The AF phase is now the only admissible phase, producing an abrupt change of the atoms without oscillations between the phases.

In particular, Ref.\cite{PRA.100.013622} reported the experimental observation of three different types of quench dynamics given by the sudden change of $q$ and with fixed magnetization of the spinor condensate. The three types of quench dynamics fits qualitatively well with the ones explained above. The metastable phase diagram in Fig.~\ref{Fig1} also helps us to predict similar quench processes given by a sudden change of the $p$ parameter instead of $q$, or by a more general sudden change involving the $(q,p)$ values.

We shall discuss now the results obtained at finite temperatures. We plot in Fig.~\ref{Fig1a} the metastable phase diagram for $T/T_0 = 0.1$ and $0.2$, which includes two new regions associated to the overlapping between the FM and AF phases (green), and a region where the three phases coexist (dark gray). The last one is associated to a free-energy $\Phi_{HF}$ with three critical points, a new scenery not seen at $T=0$. The overlapping regions among the phases are also affected by the temperature, implying that the same quench processes described above can be produced by a change of $T$ instead of $q$. For instance, let us consider a BEC in the FM  phase at $T=0.2T_0$ over the parameters $(q,p)=(-2,0.9)c_1 N$. If one cools down the BEC to nearly zero Kelvins, the FM phase is no longer an allowed  phase and it would migrate to the AF phase, the only available phase at this $(q,p)$ point. 
\begin{figure}[t!]
\begin{center}
 \scalebox{0.47}{\includegraphics{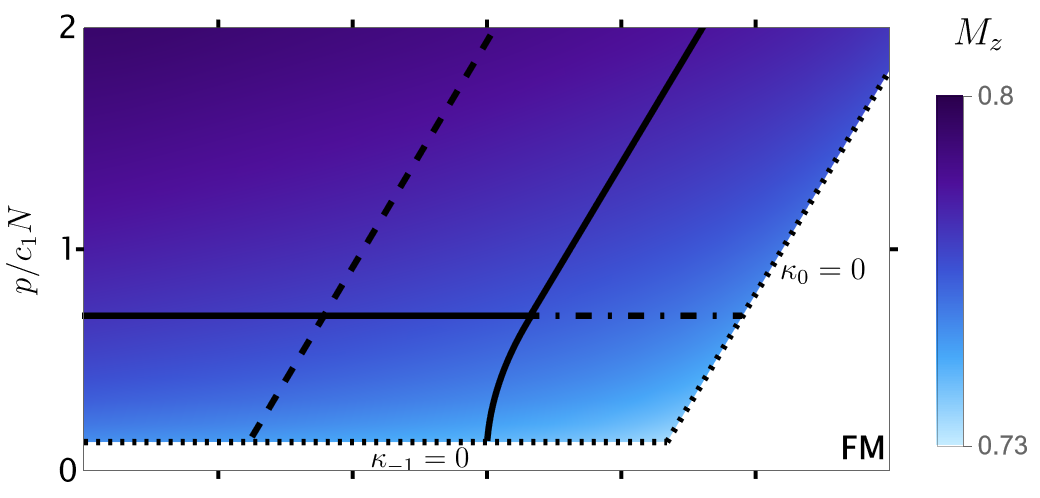}}

 \scalebox{0.47}{\includegraphics{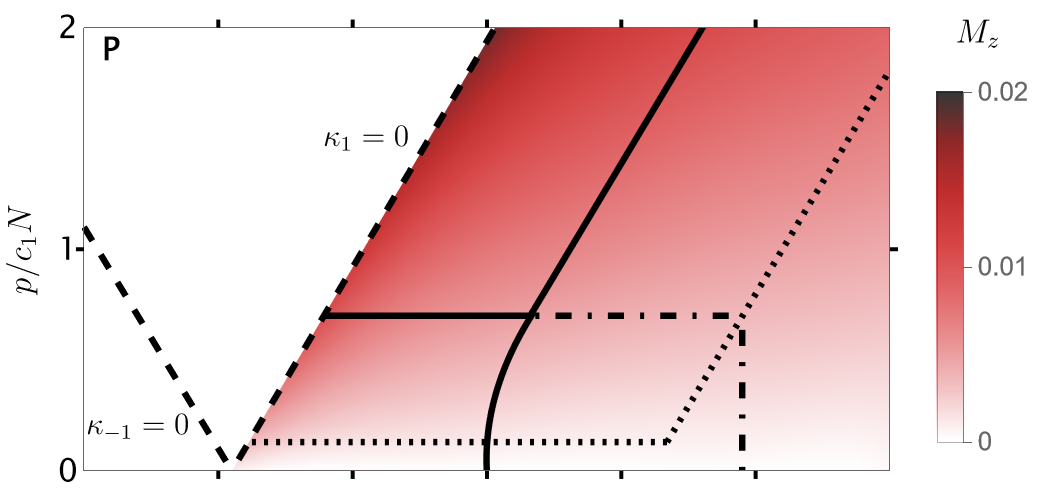}}

 \scalebox{0.47}{\includegraphics{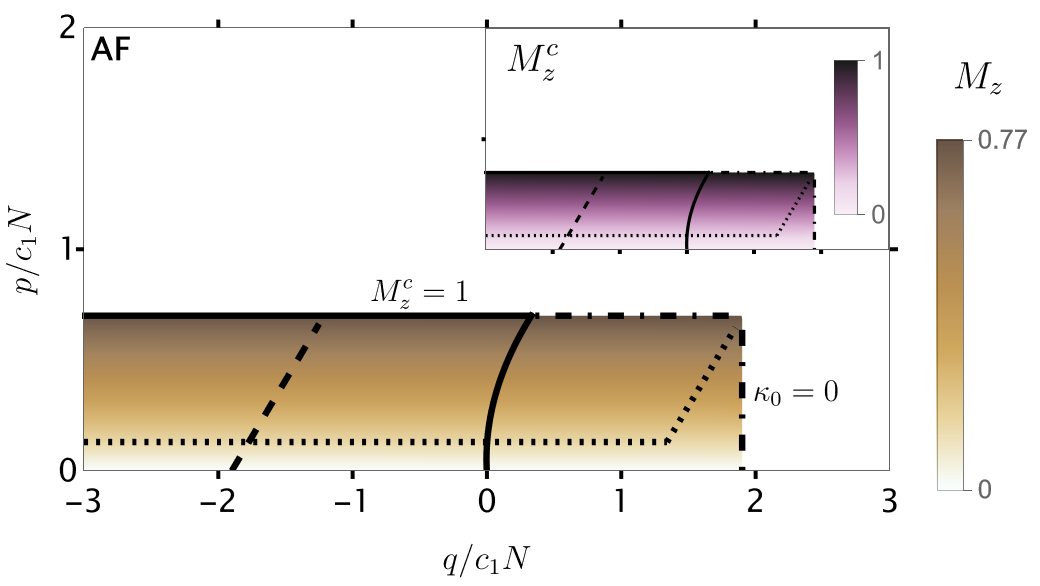}}
 \end{center}
\vspace*{-0.5cm}
\caption{\label{Fig2} (Color online) Magnetization per atom over the $z$ axis $M_z$ of the FM, P and AF phases at $T=0.2T_0$, respectively. The boundaries of each region are specified as in Fig.~\ref{Fig1}. The inset graphic in the AF phase is the magnetization per atom of the condensate fraction $M_z^c$. The color legend applies only for the admissible region of each phase.}
\end{figure}
\begin{figure*}[t!]
\begin{center}
 \scalebox{0.45}{\includegraphics{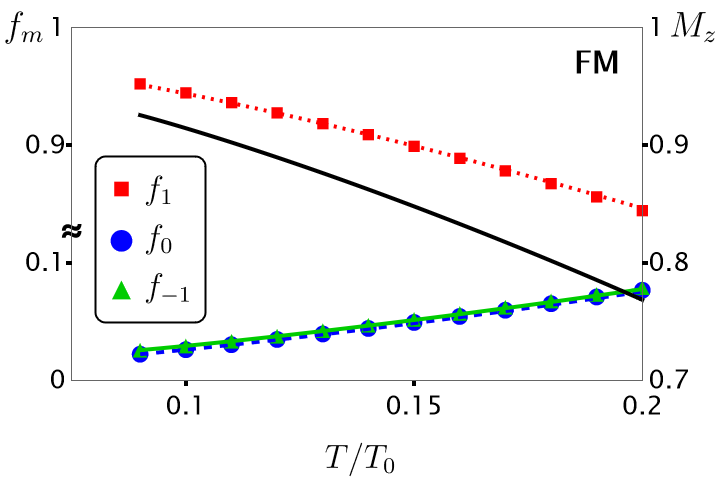}}   
 \scalebox{0.45} {\includegraphics{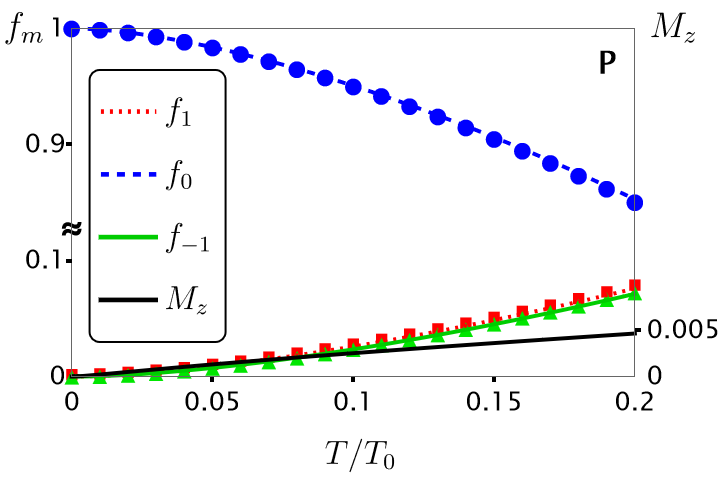}}
\hspace{-0.1cm}
 \scalebox{0.475} {\includegraphics{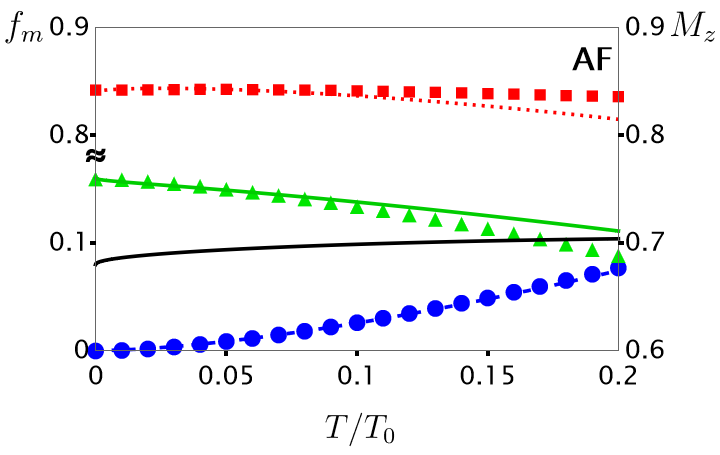}}
 \end{center}
\vspace*{-0.5cm}
\caption{\label{Fig3}  The fractions $f_m= \bra{1,m} \rho \ket{1,m}/N$ as functions of the temperature for the FM,P, and AF phases calculated numerically in the triple point at $T=0.2 T_0$, $(q,p)=(0.32 ,0.68 )c_1N $. The solid lines correspond to the fractions using the analytic expressions of $\kappa_{\nu}$ (Appendix \ref{App.B}), which  agrees well with the numerical results denoted by the squares, circles, and triangles, respectively. The black solid line corresponds to $M_z$ with the scale denoted on the right axis.}
\end{figure*}

The approach applied in this work not only leads to obtain the whole region of each phase but also the physical nature of their boundaries. The HF approximation inserts the new ingredient of the thermal atoms $\rho^{nc}$, populated with respect to the energies $\kappa_{\nu}$ plus the kinetic energy. For the atoms in the thermal cloud with $\bm{k} \approx 0$, the energy levels are given only by $\kappa_{\nu}$, which are interpreted as the additional energy to add an atom in the thermal cloud $\rho^{nc}$ instead of $\rho^c$. Therefore, $\kappa_{\nu}>0$ for all $\nu=1,2,3$, otherwise it is energetically favorable or equally to populate $\rho^{nc}$ than $\rho^c$. A phase would be forbidden as long $\ka_{\nu} \leq 0$ for any $\nu$. We list here the physical conditions of the emergence of the boundaries of each phase seen in Figs.~\ref{Fig1} and \ref{Fig1a} (see also Fig.~\ref{Fig2}):

$\bullet$ FM phase: $\rho^{nc}$ and $F_z$ share the same eigenstates $\ket{1,m}$, with $m=0,\pm1$. The two boundaries are given 
when $\kappa_{0}$ or $\kappa_{-1}=0$, where the subindex denotes the quantum number $m$.

$\bullet$ P phase: The eigenvectors of $\rho^{nc}$ are, again, the states $\ket{1,m}$, and the boundary is given by the condition $\kappa_{1}=0$. The condition $\kappa_{-1}=0$ is also plotted in Fig.~\ref{Fig2}.

$\bullet$ AF phase: One of the eigenvectors of $\rho^{nc}$ is equal to the state $\ket{1,0}$ and the other two are quantum superpositions of the states $\ket{1,\pm1}$ \cite{Kawa.Phuc.Blakie:2012} (see also Appendix \ref{App.B}). The vertical bound of the AF phase is given by $\kappa_0=0$, and the horizontal bound is a generalization of \eqref{cond.AFM1}, $M_z^c \equiv \Tr(\rho^c F_z)/ N^c =1$. 

The system of three equations of the $\kappa_{\nu}$ energies for each phase leads us to analytic approximations of the bounds in terms of the $(q,p,T)$ variables for $T < T_0$ (Appendix \ref{App.B}), 
\begin{align}
& (FM) \, \ka_0 = p-q + F_0(T) , \,\,\, 
\ka_{-1} = 2(p-c_1 N) + F_{-1}(T) \, ,
\nonumber
\\
\label{bounds.eqs}
& (P) \quad \ka_{\pm 1} = q \mp p+c_1 N +  G_{\pm 1}(T) \, ,
\\
& (AF) \, \ka_0 = c_1 N -q + H_0(T)
\, , \quad
 p= c_1 N  +H'(T) \, ,
\nonumber
\end{align}
where the last equation is given by the condition $M^c_z =1$. The functions $F_{\nu},G_{\nu}, H_{0}$, and $H'$ depend only on the temperature, and all go identically to zero when $T=0$ (Appendix \ref{App.B}). The approximations are valid up to $\mathcal{O}(k_1^2)$ with $k_1 = c_1 N/k_B T_0$, and they agree well with the numerical results obtained in the interval of $T/T_0 \in [0,0.2]$. The equations of the boundaries of each phase are deduced in Appendix \ref{App.B}. One can observe that the equations \eqref{bounds.eqs} are linear with respect to the parameters $p$ and $q$. Hence, the analytic approximation of the bounds are straight lines on the $(q,p)$-space such that its slope remains invariant but its position depends on $T$ for $T < T_0$. The regions of the phases increase along all the boundaries as we increase the temperature (see Appendix \ref{App.B}), except along the horizontal bound of the AF phase.
\subsection{Physical properties}

The BEC phases can be distinguished among each other by their physical properties. In Fig.~\ref{Fig2}, we plot the magnetization per atom $M_z \equiv \Tr(\rho F_z)/N$ of each phase in its allowed region at $T=0.2T_0$. The color density of each plot is normalized differently, and we denote the boundaries as we did in Fig.~\ref{Fig1}. The most of the magnetization of each phase arises from the condensate fraction $\rho^c$, while the small deviations can be understood by the physical origin of the boundaries:

$\bullet$ FM phase: Its magnetization  $M_z$ decreases in the $(q,p)$ values close to the boundaries. This is true because as $\kappa_{m}$ decreases, more atoms are populated to the state $\ket{1,m}$, which it has zero or negative magnetization for $m=0,-1$, respectively.

$\bullet$ P phase: The magnetization increases for the $(q,p)$ values adjacent to the line $\kappa_1=0$ and also further away to the $\kappa_{-1}=0$ condition. The BEC has null magnetization in the line $p=0$ because it is equidistant to both conditions $\kappa_{\pm1}=0$ \eqref{bounds.eqs}.

$\bullet$ AF phase: Similar as for the $T=0$ case, the maximum value of $p$ for the AF phase is when $M_z^c=1$, \ie, when the AF phase is identical to the FM phase. $M_z$ decrease as $p$ decreases, with minimum $M_z=0$ for $p=0$. Also, the numerical calculations reveal that $M_z$ is independent of the $q$ parameter, which is also confirmed with the analytic approximations for $T < T_0$ (Appendix \ref{App.C}). The inset plot of the AF phase shows the condensate magnetization $M^c_z$. By comparing both figures, one can deduce that the non-condensate fraction $\rho^{nc}$ would play against the magnetization of the condensation fraction, as it is discussed in \cite{Kawa.Phuc.Blakie:2012}.
\begin{figure}[t!]
\begin{center}
 \scalebox{0.55}{\includegraphics{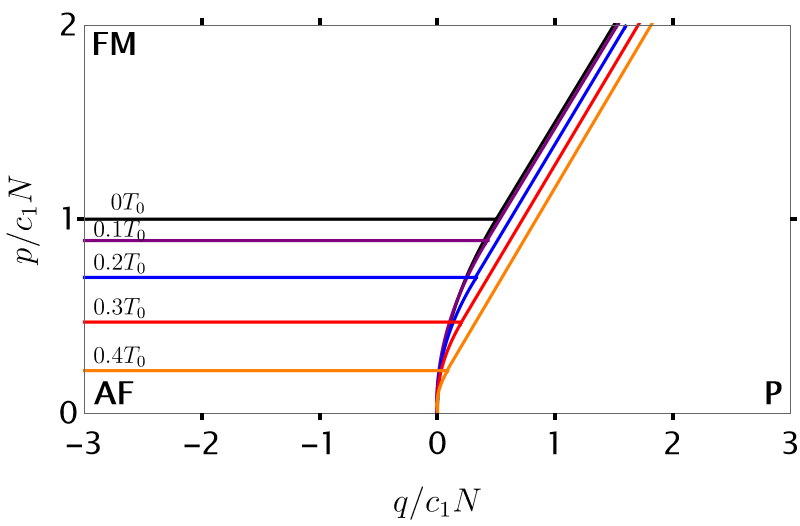}}
 \end{center}
\vspace*{-0.5cm}
\caption{\label{Fig4} Temperature dependence of the phase diagram (solid line) for $T/T_0=0,0.1,\dots,0.4$. The FM-AF boundary as a function of the temperature was described in \cite{Kawa.Phuc.Blakie:2012}. The behavior of the FM-P boundary is explained by the difference among the HF potentials of the phases \eqref{HFpot.diff}.}
\end{figure}

Another way to distinguish the phases is by their population fractions of the $\ket{1,m}$ states, which can be monitored experimentally by absorption images \cite{Jacob:12,PhysRevLett.119.050404}. In Fig.~\ref{Fig3}, we plot the fractions $f_m \equiv \bra{1,m} \rho \ket{1,m}/N$ versus the temperature for the $(q,p)$-values of the triple-point at $T=0.2T_0$,  $(0.32,0.68)c_1 N$. The P and AF phases exist for the temperatures $T/T_0 \in [0,0.2]$. On the other hand, the FM phase only exists at the temperature interval $T/T_0 \in [0.09,0.2]$. For the FM and P phases, their respective fractions begin to decrease as one increases the temperature, while the other two projections are populated equally in the thermal cloud. The AF phase increases (decreases) the fraction $f_0$ ($f_{\pm 1}$) as one increases the temperature. The change of $f_1$ from $T=0$ to $T=0.2T_0$ is more notorious as one reduces the $p$ parameter (Appendix \ref{App.D}). This result provide us a way to distinguish the AF and FM phases by comparing the evolution of the $f_1$ fraction with respect to the temperature. Note that  the change $f_1$ from $T=0$ to $T=0.2T_0$ is at least one order of magnitude greater for the FM phase as for the AF phase (see Fig.~\ref{Fig3} and Appendix \ref{App.C}). The black lines of Fig.~\ref{Fig3} correspond to $M_z = f_1 - f_{-1}$. The magnetization of FM phase is screened by the thermal atoms. On the other side, the magnetic sublevels $m=1$ ($m=0$) in $\rho^{nc}$ are more populated than the other magnetic sublevels in the P (AF) phase. Consequently, $M_z$ increases with respect to the temperature.
\subsection{Phase-transitions boundaries}
To end this section, we plot the phase diagrams for different temperatures in Fig.~\ref{Fig4}, where the ground states minimize the thermodynamic potential $\Phi_{HF}$. The FM phase reaches lower $p$-values as the temperature is increased, as it is mentioned in \cite{Kawa.Phuc.Blakie:2012}. Another feature we reveal here is that the FM phase also increases to the positive interval of the $q$ parameter. This can be better understood as we compare the thermodynamic potentials of the FM and P phases using the analytic expression of the energies $\kappa_{\nu}$ (Appendix \ref{App.B}), which leads to
\begin{equation}
\label{HFpot.diff}
\Phi_{HF}^{(FM)}-
\Phi_{HF}^{(P)} = \frac{N g}{2}\left( 2(q-p)+c_1 N g\right) + \mathcal{O}(k_1^2) \, ,
\end{equation}
where $g= g(T)= 1- 3 \zeta(3/2)/ \lambda_{dB}^3N$, with $\zeta(z)$ the Riemann zeta function. The previous equation predicts the linear behavior of the FM-P boundary in the $(q,p)$ parameters. The intersection between the AF-FM and FM-P boundaries is the triple-point of the phase diagram, and its position decreases on the $p$ and $q$ parameters as the temperature $T$ is increased (see Fig.~\ref{Fig4}).
\section{Conclusions} 
\label{sec.Con}
We have shown that a minimal many-body HF theory that fully accounts for the Hamiltonian and order parameter symmetries of a spin-1 antiferromagnetic BEC, allows us to describe the presence of regions where metastable phases could arise at zero and finite temperatures. The metastable spin-phase diagram provides a useful and complementary way to understand the different types of quench dynamics observed in experiments \cite{PRA.100.013622}, among other phenomena. 
The spin phases can be easily distinguished in laboratory by their physical properties, \eg, its magnetization or atom fractions by Stern-Gerlach spin separation \cite{PhysRevA.99.023606, jimenez2019spontaneous}. In particular, we enlightened two unique properties of the AF phase: its magnetization is independent of the  quadratic Zeeman coupling factor $q$, and the atom fraction $f_1$ of the condensate remains basically invariant for sufficiently high values of $p$ as the temperature is increased. Finally, we found a significant shift of the FM-P boundary with temperature, and derived an analytical expression for its behavior valid at low temperatures. 

This work opens up a number of routes that can be explored within the framework presented here for the study of spin-1 antiferromagnetic BEC. Indeed, the approach is quite general, as it can be straightforwardly applied to spinor condensates with different spin-dependent interactions and/or higher internal spin values. We should remark that while the ground state configuration of any spinor condensate is understood to be stable by definition, the emergent metastable states could be unstable under some weak perturbations or instabilities. Even tough, the approach is  still suitable and it can be formally extended, \eg, through  the aid of the Hessian of the thermodynamic potential to characterize the instabilities of the metastable phases.
\begin{acknowledgments}
E.S.-E. would like to acknowledge support from a postdoctoral fellowship of CONACyT. F.M. acknowledges the support of DGAPA-UNAM through the project PAPIIT No. IN113920.
\end{acknowledgments}
\bibliographystyle{apsrev4-2}\bibliography{refs_metastable}
%
\onecolumngrid
\appendix
\section{Expressions of \texorpdfstring{$E_s$}{Es} and \texorpdfstring{$S_{HF}$}{SHF}}
\label{App.A}
Here we derive the spatial energy $E_s$ in Eq. \eqref{HF.energy} and the spatial entropy $S_{HF}$ of the Hartree-Fock theory, which can be written in terms of the $\ka_{\nu}$ eigenenergies
\begin{align}
\rho^{nc}_{ij} = \sum_{\bm{k},\nu}  
n_{\la}
\xi^{\nu}_i e^{i \bm{k} \cdot \bm{r}} 
\left( \xi^{\nu}_j e^{i \bm{k} \cdot \bm{r}} \right)^{*}
= \sum_{\nu} \xi^{\nu}_i \xi^{\nu*}_j (2\pi)^{-3} (4 \pi) \int_0^{\infty} k^2 \left( z_{\nu}^{-1} e^{\frac{\beta \hbar^2 k^2}{2M}} -1 \right)^{-1} \diff k
\end{align}
where $z_{\nu} = e^{-\beta \ka_{\nu}}$. By a change of variable $x= \beta \hbar^2 k^2/2M$, we deduce that Eq. \eqref{Poly.sta}
\begin{align}
\rho^{nc}_{ij} =  \sum_{\nu} \xi^{\nu}_i \xi^{\nu*}_j \frac{4\pi\sqrt{2} (Mk_BT)^{3/2}}{h^3} \int_0^{\infty}
x^{1/2} \left( z_{\nu}^{-1} e^x-1 \right)^{-1} \diff x 
= \sum_{\nu} \xi^{\nu}_i \xi^{\nu*}_j \frac{Li_{3/2} (z_v)}{\la_{dB}^3}
\, ,
\end{align} 

\noindent with $\la_{dB} = h/(2\pi Mk_B T)^{1/2}$ the thermal de Broglie wavelength and $Li_j (z)$ the polylogarithm function. Analogously, the spatial energy $E_{s}$ is equal to 
\begin{equation}
E_{s} = \frac{\hbar^2}{2M} \sum_{\bm{k},\nu,i}  \xi^{\nu}_i e^{i \bm{k} \cdot \bm{r}} \left( \xi^{\nu}_i e^{i \bm{k} \cdot \bm{r}} \right)^{*} n_{\la} k^2  = \sum_{\nu} \frac{3 Li_{5/2}\left(z_{\nu} \right)}{2\beta \la_{dB}^3} \, .
\end{equation}
Finally, we use the same limit of the wavenumber for the HF entropy \cite{blaizot1986quantum}
\begin{equation}
S_{HF} = -k_B \sum_{\la} n_{\la} \ln n_{\la} - (1+ n_{\la}) \ln (1+ n_{\la} ) 
= \frac{k_B}{\la_{dB}^3} \sum_{\nu} \frac{5}{2} Li_{5/2}(z_{\nu}) - Li_{3/2} (z_{\nu}) \ln z_{\nu} \, .
\end{equation}
\section{Equations for the \texorpdfstring{$\kappa_{\nu}$}{knu} energies and their analytic approximations}
\label{App.B}
For the equations of the finite temperature $T\neq 0$ case, we use a similar notation as in \cite{Kawa.Phuc.Blakie:2012}, but we use the sort of the subindexes of  $\rho^c$ and $\rho^{nc}$ as in \cite{blaizot1986quantum}. The condensate fraction $\rho^c = N^c \bm{\Phi} \bm{\Phi}^{\dagger}$ is obtained by the GP equations $\delta E_{HF}/\delta \phi^*_m =0 $, which are equal to
\begin{align}
\label{HF.GPE}
\mu \bm{\Phi} = L \bm{\Phi} \, , \quad L = & -p F_z + q F_z^2+ c_0 \big( N \mathds{1}_3  + \rho^{nc} \big)
 + c_1  \sum_{\al} \Big\{ \Tr \left[F_{\al} \rho \right] F_{\al} + F_{\al} \rho^{nc} F_{\al} \Big\} \, ,
\end{align}
where $\al = x,y$ and $z$. On the other hand, the non-condensate Hamiltonian $A_{ij} = \delta E_{HF} / \delta \rho_{ji} $ has the following expression
\begin{align}
\label{HF.nc}
A = & L -\mu \mathds{1}_3 -c_0 \rho^c + c_1 \sum_{\al} F_{\al} \rho^{c} F_{\al} 
=-\mu \mathds{1}_3 -p F_z + q F_z^2 + c_0 \big( N \mathds{1}_3 + \rho \big)
 + c_1 \sum_{\al} \Big\{
 \Tr \left[ \rho F_{\al} \right] F_{\al} + F_{\al} \rho F_{\al} \Big\} \, .
\end{align}
To simplify the equations and calculations, we scale the following variables
\begin{equation}
\label{Sca.Eqs}
\left( 
\begin{array}{c}
\bar{p} \\ \bar{q} \\ \bar{L} \\ \bar{\mu} \\ \bar{A} \\ \bar{\kappa}_{\nu}
\end{array}
 \right) = \left( 
\begin{array}{c}
p \\ q \\ L \\ \mu \\ A \\ \kappa_{\nu}
\end{array}
 \right) / |c_1| N \, , \quad
\left( 
\begin{array}{c}
\bar{N}^{c} \\ \bar{N}^{nc} 
\\ \bar{\rho}^{c} \\ \bar{\rho}^{nc} \\ \bar{\rho}
\end{array}
 \right) = \left( 
\begin{array}{c}
N^c \\ N^{nc} \\ \rho^{c} \\ \rho^{nc} \\ \rho
\end{array}
 \right) / N \, , \quad
\bar{c}_0 = \frac{c_0}{|c_1|} \, , \quad 
 \quad 
\bar{T} = \frac{T}{T_0} \, .
\end{equation}
The condensate and non-condensate fractions satisfy that $\bar{N}^c+\bar{N}^{nc}=1$.  $\bar{\rho}^{nc}$ is written as 
\begin{equation}
\label{rhonc.bar}
\bar{\rho}^{nc}_{ij} = \sum_{\nu} \Lambda_{\nu} \xi^{\nu}_i \xi^{\nu*}_j 
\, , \quad \Lambda_{\nu} = \frac{Li_{3/2}\left(e^{-z_{\nu}} \right)}{N\la_{dB}^3} \, ,
\end{equation}
with $\bar{N}^{nc}= \sum_{\nu} \Lambda_{\nu}$ and
\begin{equation}
z_{\nu}=  \left( \frac{c_1 N}{k_B T_0} \right) \left( \frac{\bar{\kappa}_{\nu}}{\bar{T}} \right) = k_1 \frac{\bar{\kappa}_{\nu}}{\bar{T}} \, .
\end{equation}
In the following, we will work with the scaled variables and we will suppress the bar symbol in each term. We also define $\eta= c_1 / |c_1|$ to express our results for any type of interaction: ferromagnetic ($\eta=-1$), antiferromagnetic ($\eta=1$), or without spin-dependent interactions ($\eta=0$). The equations \eqref{HF.GPE} and \eqref{HF.nc} are reduced to
\begin{align}
\mu \bm{\Phi} = & L \bm{\Phi} \, , \quad L = -p F_z + q F_z^2+ c_0 \big( \mathds{1}_3  + \rho^{nc} \big)
 + \eta  \sum_{\al} \Big\{ \Tr \left[F_{\al} \rho \right] F_{\al} + F_{\al} \rho^{nc} F_{\al} \Big\} \, ,
\label{GPE.eq}
 \\
 A= & -\mu \mathds{1}_3 -p F_z + q F_z^2 + c_0 \big( \mathds{1}_3 + \rho \big)
 + \eta \sum_{\al} \Big\{
 \Tr \left[ \rho F_{\al} \right] F_{\al} + F_{\al} \rho F_{\al} \Big\} \, .
\label{Eq.norm}
\end{align} 
Now, we will deduce the system of equations for $\ka_{\nu}$ of each phase. The numerical results exposed through this work were obtained by solving the equations \eqref{kappa.FM}, Eqs. \eqref{kappa.P}, and the Eqs. \eqref{kappa0.AF}-\eqref{Ener2.AF} for the FM, P, and AF phase, respectively. We also derive the analytic approximations for $\ka_{\nu}$ for each phase. The analytical approximations agree well with the numerical results in the low-temperature regime (see Fig.~\ref{Fig3}).
\subsection{FM phase}
The ferromagnetic phase $\bm{\Phi}=(1,0,0)^{\text{T}}$, after it is inserted in the GP equations \eqref{GPE.eq}, yields that the chemical potential is equal to 
\begin{equation}
\mu = -p + q + c_0 \left( \La_1+1 \right) + \eta \left(1+\La_1 - 2\La_{-1} \right) \, .
\end{equation}
By a direct calculation, one obtains that the eigenvectors of $A$ are given by the eigenstates of the operator $F_z$, $\ket{1,m}$ with $m=-1,0,1$. Another way to arrive to this result is by symmetry arguments. The FM state and the Hamiltonian $\hat{H}$ \eqref{Full.Ham} have the same symmetry group $SO(2)$, associated to rotations about the $z$ axis. Hence, $\rho^{nc}$ must have the same symmetry group (See Sec.~8.4 of \cite{blaizot1986quantum}). Consequently, $\rho^{nc}$ must commute with $F_z$, and hence share the same eigenvectors $\ket{1,m}$. A useful visual way to infer the rotational symmetries over the quantum states, which allows to simplify the degrees of freedom, is through the stellar Majorana representation for pure \cite{Maj.Rep} and mixed states \cite{Ser.Bra:20}.
The eigenvalues of $A$ \eqref{Eq.norm} are described by
\begin{align}
\kappa_1^{(FM)} = & \left(c_0 + \eta \right) \left( 1- \Lambda_1 - \Lambda_0 - \Lambda_{-1} \right)  \, ,
\nonumber
\\
\kappa_0^{(FM)} = & p-q -\left( c_0 + \eta \right) \Lambda_1 + \left( c_0 - \eta \right) \Lambda_0 + 2 \eta \Lambda_{-1} \, ,
\label{kappa.FM}
\\
\kappa_{-1}^{(FM)} = & 2(p-\eta) -\left( c_0 + \eta \right) \Lambda_1 + 2 \eta \Lambda_0 + \left( c_0 + 5\eta \right) \Lambda_{-1} \, .
\nonumber
\end{align}
Here, the superscript index denotes the quantum phase. Once we write the \rhs~of the equations as functions of the energies, $\kappa_{\nu}^{(FM)} = K_{\nu}^{(FM)}\left( \kappa_1^{(FM)} , \kappa_0^{(FM)} , \kappa_1^{(FM)} \right) $, the resulting algebraic-transcendental equations cannot be solved analytically. However, we can obtain useful analytic expressions through some approximations. The first approximation is given by
\begin{equation}
\label{approx.Li}
 \Lambda_{\nu} = \frac{Li_{3/2}\left(e^{-z_{\nu}} \right)}{N\la_{dB}^3} 
\approx
\frac{1}{N\la_{dB}^3} \left( \zeta\left( \frac{3}{2} \right) - 2\sqrt{\pi z_{\nu}} - \zeta\left( \frac{1}{2} \right) z_{\nu} + \mathcal{O}\left( z_{\nu}^{2}\right) \right) \, , \quad z_{\nu} = \frac{k_1 \ka_{\nu}}{T} \, ,
\end{equation}
which is valid for our case because we are interested in the qualitative behavior around $ k_1 \kappa_{\mu} /T \approx 0 $ where $k_1 = c_1N/k_B T_0=1.75 (10^{-3})$. Here $\zeta(z)$ is the Riemann zeta function. We expand the analytic approximations with respect to $k_1$ to expose compact expressions. In addition, the numerical results tell us that $\kappa_1 > \kappa_0, \kappa_{-1}$. Then, we can also assume on the \rhs~of \eqref{kappa.FM} that 
\begin{align}
\kappa_{\nu}^{(FM)} \approx & K_{\nu}^{(FM)}\left( \kappa_1^{(FM)} , 0 , 0 \right) \, .
\end{align}
Thus, the equations \eqref{kappa.FM} can now be solved analytically leading to
\begin{align}
\kappa_1^{(FM)} 
= & g(c_0+\eta)-2k_2 T \left( \pi g k_1  (c_0+\eta)^3 \right)^{1/2} + k_1 k_2 T^{1/2} (c_0+\eta)^2 \left[g \zeta\left( \frac{1}{2} \right) + 2\pi k_2 T^{3/2}  \right] + \mathcal{O}(k_1^2) \, ,
\nonumber
\\
\kappa_0^{(FM)}= & p-q+2k_2T\left(\pi g  k_1(c_0+\eta)^{3} \right)^{1/2}+k_1 k_2 T^{1/2} (c_0+\eta)^2 \left[ g \zeta\left( \frac{1}{2} \right) -2 \pi k_2T^{3/2} \right] + \mathcal{O}(k_1^2) \, ,
\label{FMkappa.aprox}
\\
\kappa_{-1}^{(FM)}= & 2(p- \eta g ) +2 k_2 T \left( \pi g k_1 (c_0+\eta)^{3/2} \right)^{1/2}+ k_1 k_2 T^{1/2} (c_0+\eta)^2 \left[ g
\zeta \left( \frac{1}{2} \right) -2 \pi k_2T^{3/2}  \right] + \mathcal{O}(k_1^2) \, ,
\nonumber
\end{align}
where we have defined 
\begin{equation}
g=g(T) = 1- \frac{3}{\la_{dB}^3N}\zeta \left( \frac{3}{2} \right) = 1- 3k_2 T^{3/2} \zeta \left( \frac{3}{2} \right) \, ,
\end{equation}
in which $k_2 =1/ \la_{0}^3 N$ where $\la_0$ is the de Broglie wavelength at $T=T_0$. The funcions $F_{\nu}(T)$ in \eqref{bounds.eqs} are given by the difference between the Eqs. \eqref{FMkappa.aprox} and their evaluation at $T=0$, $F_{\nu}(T)= \kappa^{(FM)}_{\nu}-\kappa^{(FM)}_{\nu}|_{T=0}$. 
\subsection{P phase}
The polar phase $\bm{\Phi}=(0,1,0)^{\text{T}}$ is solved similarly as the FM phase. In this case, the GP equations leads to
\begin{equation}
\mu = c_0 \left( 1+ \La_0 \right) + \eta \left( \La_1 + \La_{-1} \right) \, .
\end{equation}
The eigenvectors of $A$ and $\rho^{nc}$ are, again, the states $\ket{1,m}$. The exact equations of $\ka_{\nu}$ are given by
\begin{align}
\kappa_1^{(P)} = & q-p+\eta +c_0 \Lambda_1 - c_0 \Lambda_0 - 3\eta \Lambda_{-1} \, ,
\nonumber
\\
\kappa_0^{(P)} = & c_0 \left(1- \Lambda_1 - \Lambda_0 - \Lambda_{-1} \right)  \, ,
\label{kappa.P}
\\
\kappa_{-1}^{(P)} = &q+p+\eta-3\eta \Lambda_1 - c_0 \Lambda_0 + c_0\Lambda_{-1} \, .
\nonumber
\end{align}
Here $\kappa_0 > \kappa_{\pm 1}$ in the low-temperature regime. Then, one can assume 
\begin{equation}
\kappa_{\nu}^{(P)} \approx K_{\nu}^{(P)}\left( 0,\kappa_0^{(P)} , 0 \right) \, ,
\end{equation}
which gives the following approximations
\begin{align}
\kappa_{\pm1}^{(P)} 
= & \mp p +q + \eta g +2k_2 T \left( k_1 \pi c_0^3 g \right)^{1/2} +
k_1 k_2 c_0^2 T^{1/2} \left[ \zeta \left( \frac{1}{2} \right)g +2\pi k_2 T^{3/2} \right] + \mathcal{O}(k_1^2) \, ,
\nonumber
\\
\kappa_0^{(P)} 
= & c_0 g + 2k_2 T \left( k_1 \pi c_0^3 g \right)^{1/2} + 
k_1 k_2 c_0^2 T^{1/2} \left[ 
\zeta \left( \frac{1}{2} \right)g +
2 \pi k_2 T^{3/2} \right] + \mathcal{O}(k_1^2) \, .
\label{Pkappa.aprox}
\end{align}
Here, we observe that the energies $\kappa_{\pm1}^{(P)} $ only differ by the sign of $p$. The functions $G_{\nu}(T)$ of \eqref{bounds.eqs} are obtained similarly as $F_{\nu}(T)$ for the FM phase, $G_{\nu}(T)= \kappa^{(P)}_{\nu}-\kappa^{(P)}_{\nu}|_{T=0} $. 
%
%
%
%
%
\subsection{AF phase}
The family of states of the AF phase are given by the order parameter, $\bm{\Phi}= (\phi_1 , \phi_0, \phi_{-1})=(\cos \chi, 0, \sin \chi)^{\text{T}}$ with $\chi \in (0,\pi/4 ]$. For a general $\chi$, $\rho^c$ is symmetric under a rotation by $\pi$ about the $z$ axis $\Rr_z(\pi)$ and by the conjugation operator $C$, which is equivalent to a reflection over the $yz$ plane. Both operators are also symmetries of the Hamiltonian $\hat{H}$ \eqref{Full.Ham}. Hence, $\rho^{nc}$ must possess the same symmetries, implying that
\begin{equation}
\rho^{nc} = \left(
\begin{array}{c c c}
a & 0 & D \\
0 & b & 0 \\
D & 0 & c
\end{array}
\right) \, ,
\end{equation}
where $D$ must be real. The state $\ket{1,0}$ would be an eigenvector of $\rho^{nc}$ and $b= \La_0$. For the other two eigenvectors, we follow a similar analysis of the equations \eqref{Eq.norm} as in 
\cite{Kawa.Phuc.Blakie:2012}. Also, we only consider antiferromagnetic interactions $\eta=1$. The GP equations for the $\rho^c$ and $\rho^{nc}$ mentioned above implies that \cite{Kawa.Phuc.Blakie:2012}
\begin{equation}
\label{Eig.GPE}
\left( 
\begin{array}{c c}
-\tilde{p}- \tilde{\mu} & C_-D
\\
C_- D & \tilde{p}- \tilde{\mu}
\end{array}
\right) 
\left(
\begin{array}{c}
\phi_1 \\ \phi_{-1}
\end{array}
 \right) 
= 
\left(
\begin{array}{c}
0 \\ 0
\end{array}
 \right) \, ,
\end{equation}
where 
\begin{equation}
\tilde{p}= p - N^c \cos(2\chi) - \frac{(c_0+3)(a-c)}{2} \, , \quad \tilde{\mu} = \mu - \left( q+ c_0 + b + \frac{C_+(a+c)}{2} \right) \, , \quad C_{\pm} = c_0 \pm 1 \, .
\end{equation}
Let us remember that $N^c=1 - N^{nc} = 1-(a+b+c)$. The eigensystem \eqref{Eig.GPE} yields \cite{Kawa.Phuc.Blakie:2012}
\begin{align}
\tilde{\mu} =& \pm \sqrt{\tilde{p}^2 + C_-^2D^2} \, ,
\\
\tan (\chi) = & -\frac{C_- D}{\tilde{p}+ \sqrt{\tilde{p}^2+C_-^2D^2}} \, ,
\label{Eq1.x}
\end{align}
where we assume the negative value of $\tilde{\mu}$ to consider the lowest chemical potential. Eq. \eqref{Eq1.x} implies that $D$ is negative because $\chi \in [0,\pi/4]$. On the other hand, the Hamiltonian $A$ has, as we expected, the state $\ket{1,0}$ as eigenvector, with 
\begin{equation}
\label{kappa0.AF}
\kappa_0 =  c_0(1+ \La_0) + 1 - \La_0 - \mu \, .
\end{equation}
The other two eigenvectors are deduced by the reduced $2\times 2$ Hamiltonian $\tilde{A}$ that involves only the components $\ket{1,\pm 1}$
\begin{equation*}
\tilde{A} = \left( -\tilde{\mu} + \frac{C_+ N^c}{2} \right) \mathds{1}_2
+ \left(
\begin{array}{c c}
-\tilde{p} + \frac{C_+}{2}N^c \cos (2\chi) & 
C_- (N^c \cos \chi \sin \chi +  D )
\\
C_- (N^c \cos \chi \sin \chi +  D )
& 
\tilde{p} - \frac{C_+}{2} N^c \cos (2\chi)
\end{array}
\right)
\, .
\end{equation*}
The eigenvalues of $\tilde{A}$ are equal to \cite{Kawa.Phuc.Blakie:2012}
\begin{equation}
\label{Ener.AFpm}
\kappa_{\pm} = -\tilde{\mu} + \frac{C_+}{2}N^c \pm \sqrt{\left( \tilde{p} - \frac{C_+  N^c \cos(2\chi)}{2} \right)^2 + C_-^2 \left( N^c \cos \chi \sin \chi + D \right)^2 } \, .
\end{equation}
The $\kappa_{\pm}$ energies would give the expression of the eigenvalues of $\rho^{nc}$, which implies that
\begin{equation}
\label{Ener2.AF}
\frac{a+c}{2} \mp \sqrt{\left( \frac{a-c}{2}\right)^2 + D^2} = \La_{\pm} \, .
\end{equation}
The Hamiltonian $\tilde{A}$ and $\rho^{nc}$ have common eigenvectors, concluding that \cite{Kawa.Phuc.Blakie:2012}
\begin{equation}
\label{Eq2.x}
C_- \left(D + N^c\sin\chi \cos \chi \right)\frac{a-c}{2} = \left( \frac{C_+}{2}N^c \cos (2\chi) - \tilde{p} \right) D \, .
\end{equation}
For the FM and P phases, the unknowns quantities were given by the energies $\kappa_{\nu}$ and we only needed three equations \eqref{kappa.FM} and \eqref{kappa.P}, respectively. In this case, we have five unknown quantities $(\chi , D , \kappa_{\nu})$ to determine with the equations \eqref{Eq1.x}, \eqref{kappa0.AF}, \eqref{Ener2.AF} and \eqref{Eq2.x}. The variables $\chi$ and $D$ can be written in terms of the $a,b,c$ variables (and then with $\kappa_{\nu}$) with the equations \eqref{Eq1.x}, \eqref{Eq2.x}. In particular, the equation for $\chi$ is given by 
\begin{align}
\label{chi.eq}
&\cos^2 \chi = 
\\
&\frac{2 (c_0+3)N^c+\sqrt{4 (1-c_0 (c_0+4)) p (a-c)+8 (c_0 (c_0+2)-1) (a-c)^2+(c_0+1)^2 p^2}-4 (c_0+2) (a-c)+(c_0+5) p}{4(c_0+3) N^c} \, .
\nonumber
\end{align}
The resulting equations for $\chi$ and $D$ can be substituted in the equations \eqref{kappa0.AF} and \eqref{Ener2.AF} to obtain the system of three equations for the energies $\kappa_{\nu}$. We solve the final equations numerically to obtain the main results of the text.

Now, let us calculate the analytic approximations for $\kappa_{\nu}$ closer to the bounds 
\begin{equation}
\kappa_0^{(AF)}=0 \, , \quad M_z^{c}=1 \Longleftrightarrow \chi = 0 \, .
\end{equation}
First, we remark from \eqref{Eq1.x} that $\chi=0$ implies that $D=0$. Hence, we consider that $\chi\approx 0$, $D\approx 0$, and $\ka_0 \approx 0$ in \eqref{Ener.AFpm} and \eqref{Ener2.AF}, that yields
\begin{align}
 a \approx & \La_{+} \approx k_2 T^{3/2} \left( \zeta \left( \frac{3}{2} \right)   
-2 \sqrt{\frac{\pi k_1 \kappa_+}{T}} - 
\zeta\left( \frac{1}{2} \right) \frac{k_1 \kappa_+}{T} 
 \right) \, ,
 \nonumber
 \\
 c \approx & \La_- \approx k_2 T^{3/2} \left( \zeta \left( \frac{3}{2} \right)   
-2 \sqrt{\frac{\pi k_1 \kappa_-}{T}} - 
\zeta\left( \frac{1}{2} \right) \frac{k_1 \kappa_-}{T} 
 \right) \, ,
\label{Eqsac}
\\
b= & \La_0 \approx \La_0 (\kappa_0=0)= k_2 T^{3/2} \zeta \left( \frac{3}{2} \right) \, .
\nonumber
\end{align}
with 
\begin{align}
\kappa_+ \approx & (c_0+1)N^c \approx (c_0+1)\left( 1-a-c- k_2 T^{3/2} \zeta \left(\frac{3}{2} \right) \right) \, ,
\nonumber
\\
\kappa_- \approx & 2\left(p- N^c-(c_0+3)\left(\frac{a-c}{2} \right) \right) \approx
2\left( p -1 +a+c+k_2 T^{3/2} \zeta \left(\frac{3}{2} \right) \right) - (c_0+3)(a-c) \, .
\label{kappapm.approx}
\end{align}
We linearize the Eqs. \eqref{Eqsac} with respect to $a$ and $c$ and we solve them. The expressions up to first order with respect to $k_1$ are
\begin{align}
a= & \frac{ k_2 T^{3/2}\zeta\left(\frac{3}{2} \right)}{2}
-
 \frac{\left(c_0+1 \right)k_1 k_2 T^{1/2}}{2\zeta\left(\frac{3}{2} \right)} 
\left\{
\zeta\left(\frac{1}{2} \right)
\zeta\left(\frac{3}{2} \right)
g
+ 4\pi g'
\right\} + \mathcal{O}(k_1^2)
\, ,
\nonumber
\\
c= &  \frac{ k_2 T^{3/2}\zeta\left(\frac{3}{2} \right)}{2}
+
\frac{k_1 k_2 T^{1/2}}{\zeta\left(\frac{3}{2} \right)} \left\{
\zeta\left(\frac{1}{2} \right)
\zeta\left(\frac{3}{2} \right)
(g -p ) +
4 \pi \left(g' -p \right)
\right\} + \mathcal{O}(k_1^2)
\, ,
\label{AF.ac}
\\
b = & \La_0 \approx \La_0 (\kappa_0=0)= k_2 T^{3/2} \zeta \left( \frac{3}{2} \right) \, .
\nonumber
\end{align}
with $g' = 1- 2k_2 T^{3/2} \zeta(3/2)$. 
The horizontal bound $\cos \chi =1$ in Eq.~\eqref{chi.eq} gives a quadratic equation for $p$. The solution that coincides with $p=1$ at $T=0$ is, by substituting \eqref{AF.ac}, equal to
\begin{equation}
\label{cos.chi}
p= g' - \frac{k_1k_2 T^{1/2}}{(c_0+1)\zeta \left( \frac{3}{2} \right)}
\left\{
 c_0 (c_0+1) \left( 4 \pi g' + \zeta \left( \frac{1}{2} \right) \zeta \left( \frac{3}{2} \right) \right)
-
(c_0+2)(3c_0+1)k_2 T^{3/2}
\zeta\left(\frac{1}{2} \right)
\zeta\left(\frac{3}{2} \right)^2
\right\}  + \mathcal{O} \left(k_1^2 \right)\, .
\end{equation}
The expression of $H'(T)$ in Eq.~\eqref{bounds.eqs} is given by the difference between the \rhs~ of Eq. \eqref{cos.chi} and its evaluation at $T=0$. 

Finally, let us approximate the vertical bound
 $\ka_0^{(AF)}=0$. We use the Eq. \eqref{kappa0.AF} with $\tilde{\mu}\approx- \tilde{p}$ 
\begin{equation}
\ka_0 \approx (c_0- 2)\La_0 + 1  + \tilde{p} - q - (c_0+1)\frac{(a+c)}{2} 
= (c_0- 2)\La_0 + 1 + p - N^c \cos(2\chi) - \frac{(c_0+3)(a-c)}{2} - q - (c_0+1)\frac{(a+c)}{2}
\, . 
\end{equation}
We approximate $\La_0$ on the \rhs~of the equation with \eqref{approx.Li}, and we solve the resulting equation for $\ka_0$ 
\begin{align}
\ka_0^{(AF)} = & 1-q + \frac{c_0-5}{2}k_2 T^{3/2} \zeta\left( \frac{3}{2} \right)
+ \frac{k_1k_2 T^{1/2}}{(c_0+1)\zeta \left( \frac{3}{2} \right)}\left\{
(k_3 g + 4c_0 p) \zeta \left( \frac{1}{2} \right) \zeta \left( \frac{3}{2} \right) + 4\pi (k_3 g' + 4c_0 p)
\right\} + \mathcal{O} \left(k_1^2 \right) \, .
\end{align}
with $k_3=(c_0-1)(c_0^2+2c_0-1)$. 
We can observe that the condition $\ka_0^{(AF)}=0$ also depends on $p$ and then the boundary line is not a vertical line. However, $k_3 \gg 4c_0 p $ for the values that we consider in this work for $p$ and $T$. Hence, one can neglect the term $4c_0p$
\begin{equation}
\ka_0^{(AF)} \approx 1-q + \frac{c_0-5}{2}k_2 T^{3/2} \zeta\left( \frac{3}{2} \right)
+ \frac{k_1k_2 k_3 T^{1/2}}{(c_0+1)\zeta \left( \frac{3}{2} \right)}\left\{ g
 \zeta \left( \frac{1}{2} \right) \zeta \left( \frac{3}{2} \right) + 4\pi g'
\right\} + \mathcal{O} \left(k_1^2 \right) \, .
\end{equation} 
From the previous equation, one can obtain the expression of $H_0(T)$ of \eqref{bounds.eqs}. 
%
%
%
%
%
%
%

For completeness, we calculate also the analytic approximations of $\kappa_{\pm}$ by inserting Eqs. \eqref{AF.ac} in \eqref{kappapm.approx},
\begin{align}
\kappa_{+}^{(AF)} =& 
\frac{(c_0+1)}{2 \zeta \left(\frac{3}{2} \right)} \left\{ 
2\zeta \left(\frac{3}{2} \right) g' + k_1 k_2T^{1/2} \left[
4\pi ((c_0-1)g'+2p) + ((c_0-1)g+2p)\zeta\left(\frac{1}{2} \right)\zeta\left(\frac{3}{2} \right)
\right]
\right\} + \mathcal{O}(k_1^2)
\, , 
\\
\kappa_{-}^{(AF)} =& 
2(p- g') 
\\
& +  \frac{k_1 k_2 T^{1/2}}{2\zeta \left( \frac{3}{2} \right)}\left\{ 
4\pi \left( (c_0^2+4c_0+11)g' - 2p(c_0+5) \right) 
+ \left( (c_0^2+4c_0+11)g - 2p(c_0+5) \right) \zeta \left( \frac{1}{2} \right) \zeta \left( \frac{3}{2} \right) 
\right\} + \mathcal{O}(k_1^2) \, .
\nonumber
\end{align}
We can deduce from the previous equations that the analog functions $H_{\pm}$ for the $\kappa_{\pm}$ energies are dependent of the variables $p$ and $T$. The system of equations for the AF phase is more complicated due to the calculation of two eigenvectors of $A$. Then, our approximations would be less accurate as in the previous phases (see Fig.~\ref{Fig3}). Even tough, the approximations are acceptable with the numerical calculations for $T \leq 0.2 T_0$.
\section{Analytic approach of the total magnetization in the AF phase}
\label{App.C}
Fig.~\ref{Fig2} indicates that the magnetization per atom  $M_z$ of the AF phase is independent of the $q$ variable. We can prove the previous statement for low temperatures by calculating $M_z$ with the approximate expressions found in the previous appendix
\begin{align}
M_z = \Tr (\rho F_z) = & \cos (2\chi)N^c + a-c 
\\
\approx &  p + \frac{c_0 k_1 k_2 T^{1/2}}{(c_0+1)\zeta \left( \frac{3}{2}\right)}\left\{
\zeta \left( \frac{1}{2}\right) 
\zeta \left( \frac{3}{2}\right) 
\left( (c_0+3)g - 2p \right)
+ 4\pi ((c_0+3)g' -2p)
\right\} + \mathcal{O}(k_1^2) \, .
\nonumber
\end{align}
\section{Change of the fraction \texorpdfstring{$f_1$}{f1} for the AF phase with respect to the temperature}
\label{App.D}
\begin{figure}[t!]
\begin{center}
 \scalebox{0.55}{\includegraphics{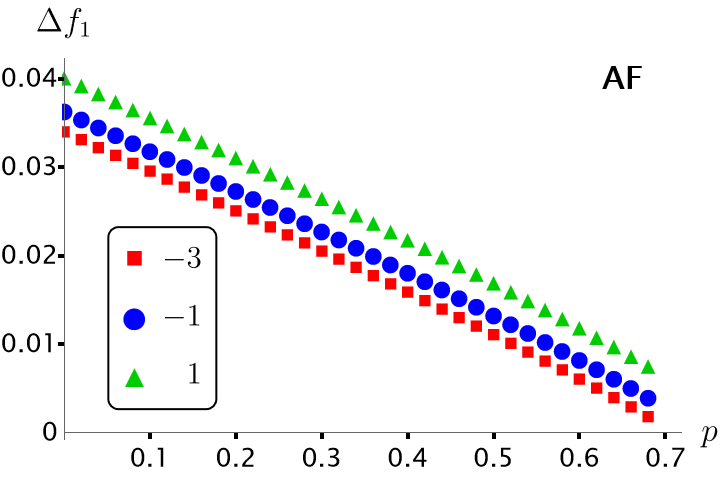}
 }
 \end{center}
\vspace*{-0.5cm}
\caption{\label{Fig4A} The change of the fraction $f_1$ for the AF phase from $T=0$ to $T=0.2$ $\Delta f_1$ \eqref{Del.f1} versus $p$ for $q=-3,-1,1$, respectively. $\Delta f_1$ decreases as one increases $p$. $\Delta f_1$ for $p=0.68$ is at most 0.01.}
\end{figure}
In this appendix, we summary the numerical results regarding the change of the fraction $f_1 = \bra{1,1} \rho \ket{1,1}$ for the AF phase with respect to the temperature. We define
\begin{equation}
\label{Del.f1}
\Delta f_1 \equiv \left. f_1 \right|_{T=0.2} - \left. f_1 \right|_{T=0}
\, ,
\end{equation} 
where we are using the scaled variables.
$\Delta f_1$ is a function of the variables $(p,q)$ and it is well-defined only when the AF phase exists for $T=0$ and $T=0.2$, \ie, for $q \in [-3,1]$ and $p\in [0,0.68]$. We plot in Fig.~\ref{Fig4A} $\Delta f_1$ versus the $p$ variable for $q=-3,-1$ and $1$, respectively. One can conclude that the fraction $f_1$ is almost invariant for $p=0.68$ in the interval of temperatures $T\in [0,0.2]$, where $\Delta f_1$ is at most equal to $0.01$. 
\end{document}